\documentclass[aps,prl,reprint,amsmath,amssymb,nofootinbib,longbibliography,raggedbottom,floatfix]{revtex4-2}
\ifdefined\pdfminorversion\pdfminorversion=7\fi
\usepackage[T1]{fontenc}
\usepackage{lmodern}
\usepackage{graphicx,booktabs,xcolor,hyperref,microtype,placeins,mathrsfs}
\graphicspath{{figures/}}
\definecolor{inkblue}{HTML}{24577A}
\hypersetup{colorlinks=true,linkcolor=inkblue,citecolor=inkblue,urlcolor=inkblue,
 pdftitle={Multiloop Unitarity from Airy Zeros},pdfauthor={Laurentiu Rodina}}
\newcommand{\Tr}{\operatorname{Tr}}
\newcommand{\Res}{\operatorname{Res}}
\begin{document}
\title{Multiloop Unitarity from Airy Zeros}
\author{Laurentiu Rodina}
\affiliation{Beijing Institute of Mathematical Sciences and Applications (BIMSA), Beijing 101408, China}
\date{\today}
\begin{abstract}
Hidden zeros are non-trivial cancellations that make amplitudes or integrands vanish on specific kinematic loci away from propagator poles. They have been shown to uniquely determine tree amplitudes and one-loop surface integrands in $\Tr(\phi^3)$ theory, without assuming factorization. Here we extend both the zeros and the uniqueness results to two, three and four loops at arbitrary multiplicity. We show that two types of zeros naturally enter the all-loop problem. First, tree-level zeros extend to all loop orders and fix relative weights within families of subgraphs attached to vacuum ``loop cores.'' We then show that Kontsevich's matrix Airy function can be used to generate a rich structure of loop level zeros that fix the relative weights of the loop cores themselves. The relative factorization and cut data required by unitarity therefore follow from the combination of tree and Airy loop zeros. These results bring within reach the emergence of unitarity to all orders in perturbation theory.
\end{abstract}
\maketitle

\section{Introduction}
Unitarity ties scattering processes together: the singularities of an amplitude must describe consistent propagation and scattering of intermediate particles. On-shell recursion and loop unitarity methods use factorization poles and cuts as inputs~\cite{BCFW,BDDK,GeneralizedUnitarity}. \emph{Hidden zeros} offer another kind of input. They are non-trivial cancellations for which an amplitude or integrand vanishes on a specific kinematic locus, while the individual diagrams remain generically nonzero and no propagator is singular~\cite{Hidden}.

Zeros also connect to exact splitting formulas, which express amplitudes as products on regular kinematic loci~\cite{SmoothSplitting,Hidden,UniversalSplitting,UniversalSplittingLong,Splits,YMNewFactorizations,YMFactorizationProof,OnShellSplittingZeros,OnShellTwoSplit,HigherDerivativeSplitting,LiRodinaSplittings,LiRodinaWangSplitting,ZhouDiagrams,UniversalDiagrammaticZeros,SplittingDiagramsExpansions,SplittingBCFW}. Related zeros and splitting relations have been studied in a broad range of amplitudes, form factors and cosmological wavefunctions~\cite{DoubleCopyZeros,ExceptionalZeros,PionZerosFactorization,ZerosExpansions,HigherDerivativeZeros,MassiveZeros,FermionicZeros,StringAllZeros,IntegerRectangles,StringKLTUniversality,CosmologicalZeros,LiRodinaCosmo,FormFactorsString,FormFactorSplitting,CEGMSplitting,CEGMNLSM,NLSMZeroRecursion,OpenStringRepresentations}. 

In $\Tr(\phi^3)$ theory, hidden zeros were shown to determine tree amplitudes and one-loop surface integrands up to normalization without assuming unitarity, making factorization a consequence of the zeros~\cite{RodinaTree,BR}. At tree level, they also determine locality within a prescribed planar pole alphabet~\cite{CombinatorialLocality}. These results are part of a broader program to derive locality and unitarity from properties such as gauge invariance, soft behavior and large-momentum scaling~\cite{ART,RodinaAdler,EFTSoft,RodinaSoft,RodinaBCFW,CarrascoRodinaUV,RodinaCHYUV}. Most of the results have focused on tree amplitudes, with the above mentioned one-loop uniqueness from hidden zero opening the possibility for beyond tree results. Establishing this emergence at all loop orders is a significant open question, which is now within reach.

\begin{figure}[t]
 \centering
 \includegraphics[width=.96\linewidth]{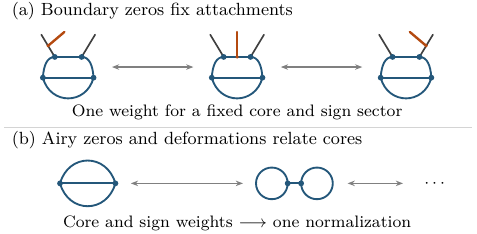}
 \caption{The two steps of reconstruction. (a) Boundary zeros compare positions of the orange external leg, fixing attachment weights for a fixed blue loop core and sign sector. (b) Airy zeros and their deformations relate the remaining core and sign weights. The arrows indicate coefficient relations. External insertions transfer the core constraints to higher multiplicity.}
 \label{fig:proof-overview}
\end{figure}

The passage from tree amplitudes to one loop already required a new kind of zero with no obvious origin. Remarkably, the one-loop zero unifies the different tree hidden zeros, which follow from it on appropriate residues~\cite{BR}. At higher loops the possible graph topologies and propagators become much more varied, and the ansatze grow rapidly. Finding such cancellation patterns, valid at every multiplicity, is therefore a substantial challenge by itself, even before attempting to prove uniqueness. To make progress at arbitrary loop order, we need a principle that guides the search for zeros strong enough to determine the integrand.

In this Letter we propose an approach that is demonstrated through four loops and can in principle be extended to arbitrary loop order. We organize the general problem around two results, illustrated in Fig.~\ref{fig:proof-overview}. First, we show that tree zeros can be extended to arbitrary loop order as \emph{boundary zeros}. These zeros then fix the relative weights of external trees attached to the same loop core. The core is the vacuum diagram obtained by removing all external legs and tree propagators. The problem then reduces to finding loop zeros that relate the different cores and their sign weights.

Second, we show that Kontsevich's matrix Airy function provides a systematic source of zeros comparing these higher loop cores~\cite{Kontsevich}. We assign a weight to each face of the graph, multiplied by its puncture sign $s=\pm$ for internal faces, and set every inverse propagator to the sum of the weights on its two sides. The resulting object is a matrix-model correlator, and summing over the $s$ gives a non-trivial zero at every multiplicity and every planar loop order $L\ge2$.\footnote{The cancellation also holds at higher genus whenever at least one internal face remains~\cite{Kontsevich}.}

However, this "Airy zero" alone is not sufficient to relate all core weights, but remarkably, it can be used to generate further families of loop zeros, by simple deformations of the propagator assignments. With locality and cubic power counting as inputs, this gives a common reconstruction argument for the two-, three- and four-loop integrands at arbitrary multiplicity. 

Together, these zeros therefore determine the factorization and cut relations required by unitarity, up to normalization. Our approach also opens a clear path to proving uniqueness at any loop order. At each fixed loop order, the boundary zero argument reduces the all-multiplicity question to a finite problem. Finding sufficiently constraining deformations of the Airy zero would establish the emergence of unitarity to all orders in perturbation theory.

\section{The local ansatz}\label{sec:local-ansatz}
\begin{figure}[!t]
 \centering
 \includegraphics[width=\linewidth]{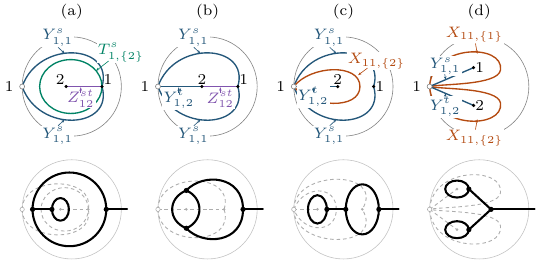}
 \caption{Two-loop one-point triangulations and their dual cubic graphs, up to puncture exchange $1\leftrightarrow2$. The curve labels distinguish boundary--puncture rays $Y$, puncture connections $Z$, returning curves $T$, and boundary curves $X$. External-tree channels $D$ first occur at two points. Restoring the exchanged partners gives eight graphs and seven projected products, with multiplicity two for (d). These one-point objects are auxiliary surface functions.}
 \label{fig:two}
\end{figure}
We define the integrands through the duality between cubic graphs and triangulations underlying the surface construction~\cite{Counting,AllMultiplicity,ScalarScaffold,TropicalLagrangians,Surface,InverseKLTSurface}. An $n$-particle, $L$-loop planar cubic graph corresponds to a triangulated disk with $n$ boundary edges and $L$ punctures. Each internal curve crosses one Feynman diagram propagator, and each triangle contains a cubic vertex. 

We name the inverse propagators according to the curves they cross. 
\begin{itemize}
\item $D_{i,d}$ joins marks $i$ and $i+d$, with boundary indices read cyclically. Together with the intervening $d$ boundary edges it bounds a region containing no punctures, and represents an external tree. Here $2\le d\le n$.
\item $X_{ij,S}$ joins boundary marks $i,j$ with punctures on both sides. For $i<j$, the nonempty proper subset $S$ lies between the curve and the directed boundary interval from $i$ to $j$. For $i=j$, we choose the zero-length interval, so $S$ is the set inside the loop based at $i$.
\item $Y_{i,p}^s$ joins boundary mark $i$ to puncture $p$.
\item $Z_{pq}^{st}$ joins distinct punctures $p,q$, with signs $s,t$ at its endpoints.
\item $T_{p,S}^s$ starts and ends at $p$ and encloses a nonempty subset $S$ of the other punctures.
\end{itemize} 
Figure~\ref{fig:projection-examples} in the appendix illustrates these various curve types.
Each puncture carries a sign $s=\pm1$, shared by every curve ending there; in $Z_{pq}^{st}$, $s$ and $t$ are the signs at $p$ and $q$, respectively~\cite{NLSMShift,CirclesTriangles,Surface,BR}. The two-loop one-point examples are shown in Fig.~\ref{fig:two}. The full integrand sums all $2^L$ choices of these signs. The relation of these objects to usual momentum-space integrands is described in Refs.~\cite{Counting,Cut}.

Beyond one loop, curves can wind around punctures in infinitely many ways, through the action of the mapping class group (MCG). Following the surface-function construction~\cite{Cut}, we identify routes related by the label-preserving MCG. Figure~\ref{fig:channel-dictionary} in the appendix illustrates different routes assigned the same variable. In each fixed sign sector this is the planar scalar surface function $G_S=\sum_\Gamma\prod_{C\in\Gamma}x_C$ of Ref.~\cite{Cut}, evaluated at $x_C=1/Q_C$. 

We write $\mathcal I_{n,L}$ for the full integrand, as a graph sum over all puncture-sign sectors. To obtain our local cubic ansatz, we now allow each distinct denominator product to carry an unknown coefficient $c_\alpha$
\begin{equation}
 M_{n,L}=\sum_\alpha \frac{c_\alpha}{\prod_{\gamma\in\alpha} Q_\gamma},
 \qquad |\alpha|=n+3L-3.
 \label{eq:ansatz}
\end{equation}
Locality requires the curves $Q_\gamma$ in each product to belong to one triangulation $\alpha$, and cubic scalar power counting makes its coefficient a constant. The coefficients are independent in every sign sector. For simplicity, we assume reflection symmetry, assigning equal coefficients to a diagram and its mirror image (see Appendix~\ref{app:reflection} for details).

\section{Boundary zeros fix the external attachments}
\label{sec:boundary-zeros}\label{sec:tree-one-zeros}

Boundary zeros are the first zero class we introduce. These extend the familiar tree-level zeros to arbitrary loop order and arbitrary multiplicity.\footnote{The all-loop zeros of Ref.~\cite{ZhouLoopZeros} constrain loop momenta and omit scaleless terms, whereas here we use zeros of independent surface variables for integrand reconstruction.} These zeros also follow from the all-order tree--loop split of Ref.~\cite{Splits}: the integrand splits into a four-point tree amplitude times a lower-point loop integrand. Setting the tree factor to zero gives the boundary zero, with the loop factor kept regular. We will show they involve the positions of an external leg while leaving the internal loop propagators arbitrary, and thereby fix the relative weights of attachments to each loop core.

At tree level, the only variables are $X_{ij}=(p_i+p_{i+1}+\ldots+p_{j-1})^2$, with $X_{i,i}=X_{i,i+1}=0$, and particle $1$ corresponds to the boundary edge between marks $1$ and $2$. Its skinny zero compares channels beginning at these two marks,
\begin{equation}
 c_{1j}=X_{1j}+X_{2,j+1}-X_{1,j+1}-X_{2j}=0,
 \label{eq:tree-skinny-boundary}
\end{equation}
for  $j=3,\ldots,n-1$. On the above locus, the $n$-point $\Tr(\phi^3)$ amplitude vanishes~\cite{Hidden}. The above conditions can equivalently be written as
\begin{equation}
 X_{2j}-X_{1j}=\eta,
 \ \textrm{for } j=3,\ldots,n,
 \qquad \eta=-X_{13}=X_{2n}.
 \label{eq:tree-skinny-difference}
\end{equation}
At loop level, the zero selecting the same external leg $1$ imposes the analogous channel differences. Suppressing puncture signs,
\begin{equation}
 Y_{2,p}-Y_{1,p}=\eta,
 \qquad X_{2j,S}-X_{1j,S}=\eta
 \quad (j=3,\ldots,n),
 \label{eq:loop-boundary-differences}
\end{equation}
for each puncture $p$ and directed subset $S$. Curves with both endpoints at marks $1,2$ satisfy $X_{12,S}-X_{11,S}=X_{22,S}-X_{12,S}=\eta$. The neighboring tree channels satisfy
\begin{equation}
 D_{1,2}=-\eta,\qquad D_{n,2}=\eta.
 \label{eq:boundary-cap-values}
\end{equation}
All $Z,T$ remain free. Together with the remaining $D$ equations in Appendix~\ref{app:boundary}, Eq.~\eqref{eq:boundary}, these relations define the \emph{boundary zero} $\mathcal B_1$, and cyclic relabelling gives $\mathcal B_i$ for each external particle. The zero holds separately in each sign sector, and leaves all propagators non-zero.

Now we want to determine the effect of imposing zero $\mathcal B_i$ on a local ansatz $M_{n,L}$, more precisely what relations the $c_\alpha$ need to satisfy for $\mathcal B_i$ to be a zero of $M_{n,L}$. This is easily understood graphically, and the argument is similar to the purely tree level one~\cite{RodinaTree}.

We delete leg $i$ from every diagram, and group resulting identical diagrams together. Boundary zero $\mathcal B_i$ then requires all diagrams that map into the same child diagram to have equal weight. Appendix~\ref{app:boundary} gives the residue proof (and explains the low-point exceptions that will be important later).

\begin{figure}[!htbp]
 \centering
 \includegraphics[width=.94\linewidth]{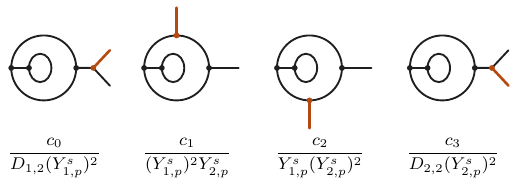}
 \caption{The four two-loop attachments in Eq.~\eqref{eq:boundary-four-attachments}. The orange leg is added to the same one-point graph. The first and last terms attach an external tree; the middle terms split the upper or lower outer-loop edge. The common internal factor $[T_{p,\{q\}}Z_{pq}]^{-1}$ is suppressed below the graphs. The boundary zero equates $c_0,c_1,c_2,c_3$.}
 \label{fig:boundary-attachments}
\end{figure}

For example, the four parents in Fig.~\ref{fig:boundary-attachments} contribute
\begin{equation}
\begin{split}
 M_{\rm att}=\frac{1}{T_{p,\{q\}}Z_{pq}}\bigg[
 &\frac{c_0}{D_{1,2}Y_{1,p}^{2}}
 +\frac{c_1}{Y_{1,p}^{2}Y_{2,p}}\\
 &+\frac{c_2}{Y_{1,p}Y_{2,p}^{2}}
 +\frac{c_3}{D_{2,2}Y_{2,p}^{2}}\bigg].
\end{split}
\label{eq:boundary-four-attachments}
\end{equation}
Their common $T,Z$ poles isolate this group. Substituting $Y_{2,p}=Y_{1,p}+\eta$, $(D_{1,2},D_{2,2})=(-\eta,\eta)$  and demanding the sum to vanish requires $c_0=c_1=c_2=c_3$.

\label{sec:core-organization}
\begin{figure}[!t]
 \centering
 \includegraphics[width=.96\linewidth]{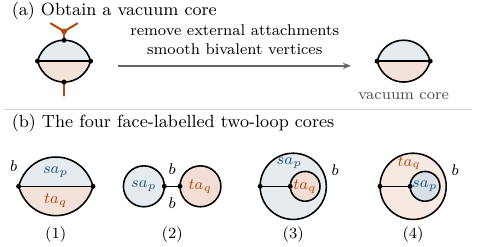}
 \caption{From attachments to the Airy vacuum sum. (a) Removing external attachments and smoothing bivalent vertices exposes a vacuum core. (b) The four two-loop cores carry one exterior-face weight $b$ and two internal-face weights. Their additive propagators give Eq.~\eqref{eq:airy-two-loop-main}. Boundary zeros fix the attachments; the Airy conditions compare the cores.}
 \label{fig:core-extraction-mountains}\label{fig:airy-faces}
\end{figure}

Imposing all boundary zeros for each external leg relates the attachments to a fixed loop core, obtained by removing all external trees and legs (Fig.~\ref{fig:core-extraction-mountains}). 

Starting at a loop-dependent multiplicity $n\ge n_L$, the boundary zeros require
\begin{equation}
 M_{n,L}=\sum_r w_{r,n} C_{r,n,L},
 \qquad n\ge n_L,
 \label{eq:boundary-core-expansion}
\end{equation}
Here $r$ labels a core and its puncture-sign sector, and $C_{r,n,L}$ sums all ways of attaching the $n$ external particles, including external trees, to that core. We choose an independent set of these complete sums as the basis.

The thresholds are $n_1=2$, $n_2=n_3=3$ and $n_4=4$, as proved in Appendix~\ref{app:boundary}. At general $L\ge2$, a simple argument bounds the finite range to check by $N_L=10L-12$, but the starting multiplicity $n_L$ can be smaller when the remaining cases are excluded sooner.  The reason Eq.(\ref{eq:boundary-core-expansion}) only holds for $n\ge n_L$ is that at lower multiplicity there are not enough boundary zeros. Each complete sum $C_{r,n,L}$ obeys every boundary zero separately, so its overall weight $w_{r,n}$ remains free. The remaining task is to find loop zeros that relate these core and sign weights.

\section{Loop zeros from the matrix Airy function}
\label{sec:airy}
\subsection{The Airy zero at every multiplicity}
We now use Kontsevich's matrix Airy function to construct zeros that relate different loop cores~\cite{Kontsevich}. Originally introduced to compute intersection numbers on moduli spaces of marked curves, this cubic matrix integral generates ribbon Feynman graphs in perturbation theory. Our vacuum loop cores appear in its planar expansion, so its exact formulas evaluate the graph sums needed here. Each face carries a weight, and the quadratic term gives an edge $\gamma$ separating faces of weights $\lambda_L$ and $\lambda_R$ the propagator
\begin{equation}
 \frac{1}{Q_\gamma}
 \xrightarrow{\text{Airy assignment}}
 \frac{1}{\lambda_L+\lambda_R}.
\end{equation}
This makes the matrix integral useful here: its identities describe cancellations among the cores on this common propagator assignment. For the four two-loop cores in Fig.~\ref{fig:airy-faces}(b), let $b$ be the exterior-face weight and $a_p,a_q$ the internal-face weights. Denote their vacuum sum by $A_3$, where the subscript counts the three labelled faces. The three propagators of the first core have inverse values $b+a_p$, $b+a_q$ and $a_p+a_q$. In the other cores, a bridge borders the same face on both sides and therefore has inverse value $2b$, $2a_p$ or $2a_q$. The four contributions give
\begin{equation}
 \begin{aligned}
 A_3(b;a_p,a_q)={}&\frac1{(b+a_p)(b+a_q)(a_p+a_q)}\\
 &+\frac1{2b(b+a_p)(b+a_q)}\\
 &+\frac1{2a_p(b+a_p)(a_p+a_q)}\\
 &+\frac1{2a_q(b+a_q)(a_p+a_q)}
 =\frac1{2ba_pa_q}.
 \end{aligned}
 \label{eq:airy-two-loop-main}
\end{equation}
The four different propagator structures have collapsed to a single product of face weights. The result changes sign when either internal-face weight changes sign. Introducing the puncture signs $\{s,t\}=\pm1$, and replacing the face weights by $sa_p,ta_q$ gives $st/(2ba_pa_q)$, so the four sign sectors cancel. This is a zero at generic face weights, where every individual propagator remains finite.

Kontsevich's formula has the same odd dependence on each internal-face weight at every planar loop order $L\ge2$~\cite{Kontsevich}. Adding external legs along the outer face splits propagators while preserving the face-sum assignment and the cancellation between puncture signs. Restoring external trees multiplies these vanishing sums by their $D^{-1}$ factors, so all $D$ can remain independent. Appendix~\ref{app:airy-insertions} gives the insertion argument.

We can therefore write the Airy zero directly for the $n$-point integrand. Give each boundary mark its own weight $b_i$ and each puncture its signed weight $sa_p$. Applying the face-sum rule to every non-tree propagator gives
\begin{equation}
 \begin{aligned}
 Y_{i,p}^s&=b_i+sa_p,&Z_{pq}^{st}&=sa_p+ta_q,\\
 X_{ij,S}&=b_i+b_j,&T_{p,S}^s&=2sa_p,
 \end{aligned}
 \label{eq:airy-locus-main}
\end{equation}
with the external-tree inverse propagators $D$ left arbitrary. We denote this assignment by $\mathcal A^\circ$ and obtain
\begin{equation}\boxed{
 \left.\mathcal I_{n,L}\right|_{\mathcal A^\circ}=0,
 \qquad L\ge2.}
 \label{eq:airy-zero}
\end{equation}
This remarkably simple Airy assignment therefore produces a non-trivial zero valid at any multiplicity and any loop order! 

However, if our goal is uniqueness, this identity is not sufficient, and so we must seek larger zero families. This can be done in a semi-systematic fashion by varying the propagators while preserving the reference cancellation. 

\subsection{Two-loop mountain deformations}
A surprisingly simple two-loop choice consists of just two one-parameter deformations. Keep $X,Z,T$ at the Airy values in Eq.~\eqref{eq:airy-locus-main}, leave every $D$ independent, and impose 
\begin{equation}
 \begin{aligned}
 Y_{i,p}^s&=b_i+s a_p+s\xi,\\
 Y_{i,q}^t&=b_i+t a_q+t \sigma\xi,
 \end{aligned}
 \label{eq:common-delta-rays}
\end{equation}
where we can choose $\sigma=+1$ or $-1$, and both assignments reduce to Airy at $\xi=0$. Appendix~\ref{app:two} gives a more general zero family.

\subsection{Three-loop deformations}
At three loops, two simple deformation rules suffice. For the first, keep $Y,X$ at their Airy values and every $D$ independent. Select one sign $\gamma_p$ at each puncture and shift 
\begin{equation}
 \begin{aligned}
 Z_{pq}^{st}&=sa_p+ta_q+
 \epsilon\,\delta_{s,\gamma_p}\delta_{t,\gamma_q},\\
 T_{p,S}^{s}&=2sa_p+\epsilon\,\delta_{s,\gamma_p}.
 \end{aligned}
 \label{eq:three-simple-zero-main}
\end{equation}
Here $\delta_{s,\gamma_p}$ is one for matching signs and zero otherwise. 

The second rule starts again from Airy and changes only one chosen return,
\begin{equation}
 T_{p,S}^s=2sa_p+\tau,
 \label{eq:three-return-releases-main}
\end{equation}
Both rules recover the Airy locus when their displacement vanishes. Choosing the signs $\gamma_p$ in the first rule, or the return $(p,S,s)$ in the second, specifies a particular zero family. Three choices of each type suffice to fix the three-loop core weights. Appendix~\ref{app:three} gives a more general zero family.

\subsection{Four-loop zeros}
At four loops, the explicit zeros are given in Appendix~\ref{app:four}.

\section{Lifting uniqueness by boundary zeros}
\label{sec:common-uniqueness}\label{sec:finite-start}

We first explicitly check uniqueness at $n_{L=2}=n_{L=3}=3$ and $n_{L=4}=4$, then extend it by a residue argument similar to Ref.~\cite{BR}.

\subsection{The finite starting cases}
\begin{figure}[!htbp]
 \centering
 \includegraphics[width=\linewidth]{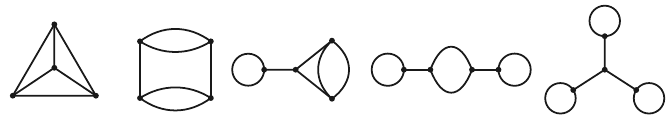}
 \caption{The five cubic vacuum topologies at three loops. The core weights also distinguish planar embeddings, the exterior face and puncture labels.}
 \label{fig:app-three-cores}
\end{figure}

At $n=n_L$, we proved the boundary zeros leave one weight for each independent core attachment sum and sign sector. We impose the loop-zero assignments introduced in the previous section on these complete $n_L$-point sums, which can be verified to leave one normalization in each case:
\begin{equation}
 M_{n_L,L}=a_{n_L,L}\mathcal I_{n_L,L},
 \label{eq:finite-start-main}
\end{equation}
for  $n_2=n_3=3$, and $n_4=4$. At two loops the two zero families reduce the sixteen weights left by the three-point boundary equations to one. At three loops, the five vacuum topologies in Fig.~\ref{fig:app-three-cores}, with their embeddings, exterior-face choices and puncture labels, give 59 independent core sums per sign sector. The six three-loop families fix their $59\times8=472$ weights. At four loops, the zero families in Appendix~\ref{app:four}, with their puncture and sign images and boundary insertions, fix the $1624\times16=25984$ four-point weights.

\subsection{Adding one external particle}
\begin{figure}[!htb]
 \centering
 \includegraphics[width=.88\linewidth]{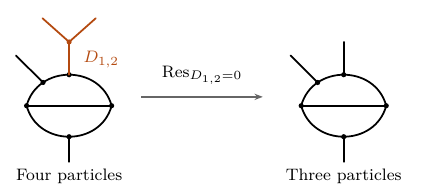}
 \caption{Removing the two-particle tree at the orange propagator preserves the core and its parent weight $w_{r,n+1}$.}
 \label{fig:uniqueness-lift}
\end{figure}

Assume uniqueness at $n$ points, with $n\ge n_L$. Taking a $D_{i,2}$ residue of an $(n+1)$-point ansatz removes a two-particle tree and leaves an $n$-point ansatz at the same loop order, as in Fig.~\ref{fig:uniqueness-lift}. Both the boundary and loop zeros descend to this residue, in the sense that they act as lower-point zeros.  Induction then gives
\begin{equation}
 R_{n,L}:=\Res_{D_{i,2}=0}M_{n+1,L}
 =a_{n+1,L}\mathcal I_{n,L},
 \label{eq:lower-residue-main}
\end{equation}
after relabelling the remaining particles. Repeating this for every $i$ fixes all  two-particle residues, and the boundary equations relate their normalizations.

These residues also determine diagrams without a $D$ propagator, which are missed by the above argument. The boundary equations have already tied every attachment to the weight of its core. Removing a two-particle tree preserves those weights, and the resulting lower-point core sums are independent for $n\ge n_L$. Equation~\eqref{eq:lower-residue-main} therefore fixes each core weight to its reference value times $a_{n+1,L}$, which determines all its attachments in the parent. Appendix~\ref{app:boundary} gives a more detailed argument.

Together with the finite checks for $n<n_L$, this proves
\begin{equation}
 \boxed{M_{n,L}=a_{n,L}\mathcal I_{n,L},\qquad \text{for } L=2,3,4,}
 \label{eq:result}
\end{equation}
so that zeros have uniquely fixed the local ansatze, up to overall normalizations. Lower multiplicities are checked directly using the zeros in Appendices~\ref{app:two}, \ref{app:three-low} and~\ref{app:four-low-points}. The accompanying Wolfram Language package checks $L=2,3$ at $n=1,2,3$, using the main-text families where sufficient and the additional low-point families in Appendix~\ref{app:families}~\cite{SM}.

\section{Factorization and loop cuts from zeros}
\label{sec:factorization}
Equation~\eqref{eq:result} fixes every relative residue and cut to that of the reference graph sum. At a simple separating pole $Q=D$ or $X$,
\begin{equation}
 \Res_{Q=0}M_{n,L}
 =a_{n,L}\mathcal I_{n_A,L_A}\mathcal I_{n_B,L_B},
 \label{eq:separating-factorization-main}
\end{equation}
with $n_A+n_B=n+2$ and $L_A+L_B=L$. A $D$ cut gives a tree factor times a loop factor; in particular, $D_{i,2}$ leaves $\mathcal I_{n-1,L}$ because the three-point tree is one. An $X$ cut separating loop regions gives two factors with positive loop order. Other separating cuts likewise give products on the two resulting surfaces.

Cutting a ray $Y_{i,p}^s$ opens puncture $p$, giving an $(n+2)$-point graph sum at loop order $L-1$, with inherited channel values and signs.  Since Eq.~\eqref{eq:result} fixes the coefficients of the graph sum, it also fixes these cut sums and their relative multiplicities. Factorization and loop cuts therefore follow from the zeros, without imposing either as a condition on the ansatz.
\section{Outlook}
We have shown that up to four loops two kinds of regular cancellation can determine the relative weights of perturbative diagrams. Boundary zeros first fix the external attachments to each loop core, and deformations of the Airy zero then relate the remaining core and sign weights. The boundary construction and the Airy zero are available at arbitrary loop order, while the general bound $N_L=10L-12$ ensures finite multiplicity checks can start the inductive step at any finite loop order. The same induction argument can therefore establish uniqueness at every higher multiplicity. This already makes higher loop proofs accessible, while further optimizations of the construction can realistically lead to an all loop-order proof.

An all-orders uniqueness theorem would show that the full hierarchy of relative factorization and cut relations follows from hidden zeros at regular kinematics.  This would establish hidden zeros as a new organizing principle for perturbative scattering. Extending the same mechanism to other theories is a next step of this program.

Understanding deformations of the Airy function directly in the matrix model could make their construction systematic. The factorization relations of surface integrands may in turn suggest new ways to organize ribbon-graph correlators and their relation to intersection theory and two-dimensional gravity~\cite{WittenGravity,Kontsevich,GrossNewman,PolynomialAverages}.

\medskip
\emph{Acknowledgments}
LR is supported by the National Natural Science Foundation
of China General Program No. 12475070 and the Beijing
Natural Science Foundation International Scientist Project
No. IS24014. AI tools were used as research and coding
assistants for exploratory calculations, checks, and
manuscript editing.

\bibliographystyle{apsrev4-2}
\bibliography{references}

@article{BCFW,
  author = {R. Britto and F. Cachazo and B. Feng and E. Witten},
  title = {{Direct Proof of the Tree-Level Scattering Amplitude Recursion Relation in Yang-Mills Theory}},
  journal = {Phys. Rev. Lett.},
  volume = {94},
  pages = {181602},
  year = {2005},
  eprint = {hep-th/0501052},
  archivePrefix = {arXiv},
  url = {https://arxiv.org/abs/hep-th/0501052}
}

@article{BDDK,
  author = {Z. Bern and L. Dixon and D. C. Dunbar and D. A. Kosower},
  title = {{One-loop $n$-point gauge theory amplitudes, unitarity and collinear limits}},
  journal = {Nucl. Phys. B},
  volume = {425},
  pages = {217--260},
  year = {1994},
  eprint = {hep-ph/9403226},
  archivePrefix = {arXiv},
  url = {https://arxiv.org/abs/hep-ph/9403226}
}

@article{GeneralizedUnitarity,
  author = {R. Britto and F. Cachazo and B. Feng},
  title = {{Generalized unitarity and one-loop amplitudes in $N=4$ super-Yang--Mills}},
  journal = {Nucl. Phys. B},
  volume = {725},
  pages = {275--305},
  year = {2005},
  eprint = {hep-th/0412103},
  archivePrefix = {arXiv},
  url = {https://arxiv.org/abs/hep-th/0412103}
}

@article{Cut,
  author = {N. Arkani-Hamed and H. Frost and G. Salvatori},
  title = {{The Cut Equation}},
  journal = {J. High Energy Phys.},
  volume = {2026},
  number = {06},
  year = {2026},
  pages = {102},
  eprint = {2412.21027},
  archivePrefix = {arXiv},
  url = {https://arxiv.org/abs/2412.21027}
}

@article{Hidden,
  author = {N. Arkani-Hamed and Q. Cao and J. Dong and C. Figueiredo and S. He},
  title = {{Hidden zeros for particle/string amplitudes and the unity of colored scalars, pions and gluons}},
  journal = {J. High Energy Phys.},
  volume = {2024},
  number = {10},
  year = {2024},
  pages = {231},
  eprint = {2312.16282},
  archivePrefix = {arXiv},
  url = {https://arxiv.org/abs/2312.16282}
}

@article{SmoothSplitting,
  author = {F. Cachazo and N. Early and Gim{\'e}nez Umbert, B.},
  title = {{Smoothly splitting amplitudes and semi-locality}},
  journal = {J. High Energy Phys.},
  volume = {2022},
  number = {08},
  year = {2022},
  pages = {252},
  eprint = {2112.14191},
  archivePrefix = {arXiv},
  url = {https://arxiv.org/abs/2112.14191}
}

@article{UniversalSplitting,
  author = {Q. Cao and J. Dong and S. He and C. Shi},
  title = {{A universal splitting of tree-level string and particle scattering amplitudes}},
  journal = {Phys. Lett. B},
  volume = {856},
  pages = {138934},
  year = {2024},
  eprint = {2403.08855},
  archivePrefix = {arXiv},
  url = {https://arxiv.org/abs/2403.08855}
}

@article{UniversalSplittingLong,
  author = {Q. Cao and J. Dong and S. He and C. Shi and F. Zhu},
  title = {{On universal splittings of tree-level particle and string scattering amplitudes}},
  journal = {J. High Energy Phys.},
  volume = {2024},
  number = {09},
  year = {2024},
  pages = {049},
  eprint = {2406.03838},
  archivePrefix = {arXiv},
  url = {https://arxiv.org/abs/2406.03838},
  note = {Erratum: J. High Energy Phys. 01 (2025) 126}
}

@article{Splits,
  author = {N. Arkani-Hamed and C. Figueiredo},
  title = {{All-order splits and multi-soft limits for particle and string amplitudes}},
  journal = {J. High Energy Phys.},
  volume = {2025},
  number = {10},
  year = {2025},
  pages = {077},
  eprint = {2405.09608},
  archivePrefix = {arXiv},
  url = {https://arxiv.org/abs/2405.09608}
}

@misc{YMNewFactorizations,
  author = {A. Guevara and Y. Zhang},
  title = {{New factorizations of Yang-Mills amplitudes}},
  year = {2024},
  eprint = {2406.08969},
  archivePrefix = {arXiv},
  url = {https://arxiv.org/abs/2406.08969}
}

@article{YMFactorizationProof,
  author = {Y. Zhang},
  title = {{On the new factorizations of Yang-Mills amplitudes}},
  journal = {J. High Energy Phys.},
  volume = {2025},
  number = {02},
  year = {2025},
  pages = {074},
  eprint = {2412.15198},
  archivePrefix = {arXiv},
  url = {https://arxiv.org/abs/2412.15198}
}

@article{OnShellSplittingZeros,
  author = {C. R. T. Jones and S. Paranjape},
  title = {{Smooth splitting and zeros from on-shell recursion}},
  journal = {J. High Energy Phys.},
  volume = {2025},
  number = {07},
  year = {2025},
  pages = {251},
  eprint = {2505.02520},
  archivePrefix = {arXiv},
  url = {https://arxiv.org/abs/2505.02520}
}

@article{OnShellTwoSplit,
  author = {T. Azevedo and H. Gomez and Lipinski Jusinskas, R.},
  title = {{On-shell representation and further instances of the 2-split behavior of amplitudes}},
  journal = {J. High Energy Phys.},
  volume = {2026},
  number = {06},
  year = {2026},
  pages = {001},
  eprint = {2512.20790},
  archivePrefix = {arXiv},
  url = {https://arxiv.org/abs/2512.20790}
}

@article{HigherDerivativeSplitting,
  author = {L. Zhang and K. Zhou},
  title = {{Generalized $2$-split for higher-derivative YM and GR amplitudes at tree-level}},
  journal = {J. High Energy Phys.},
  volume = {2026},
  number = {06},
  year = {2026},
  pages = {105},
  eprint = {2601.00297},
  archivePrefix = {arXiv},
  url = {https://arxiv.org/abs/2601.00297}
}

@misc{LiRodinaSplittings,
  author = {Y. Li and L. Rodina},
  title = {{From Hidden Zeros to Exact Splittings and New Zeros}},
  year = {2026},
  eprint = {2610.01775},
  archivePrefix = {arXiv},
  url = {https://arxiv.org/abs/2610.01775}
}

@misc{LiRodinaWangSplitting,
  author = {Y. Li and L. Rodina and Z.-Z. Wang},
  title = {{Exact Splitting of Amplitudes and One-Loop Integrands: A Classical Origin of Zeros}},
  year = {2026},
  eprint = {2611.06810},
  archivePrefix = {arXiv},
  url = {https://arxiv.org/abs/2611.06810}
}

@article{DoubleCopyZeros,
  author = {C. Bartsch and T. V. Brown and K. Kampf and U. Oktem and S. Paranjape and J. Trnka},
  title = {{Hidden amplitude zeros from the double-copy map}},
  journal = {Phys. Rev. D},
  volume = {111},
  pages = {045019},
  year = {2025},
  eprint = {2403.10594},
  archivePrefix = {arXiv},
  url = {https://arxiv.org/abs/2403.10594}
}

@article{ExceptionalZeros,
  author = {Y. Li and D. Roest and T. ter Veldhuis},
  title = {{Hidden zeros in exceptional field theories from double copy}},
  journal = {J. High Energy Phys.},
  volume = {2025},
  number = {04},
  year = {2025},
  pages = {121},
  eprint = {2403.12939},
  archivePrefix = {arXiv},
  url = {https://arxiv.org/abs/2403.12939}
}

@article{PionZerosFactorization,
  author = {Y. Li and T. Wang and T. Brauner and D. Roest},
  title = {{Diagrammatic Derivation of Hidden Zeros and Exact Factorization of Pion Scattering Amplitudes}},
  journal = {Phys. Rev. Lett.},
  volume = {135},
  pages = {181601},
  year = {2025},
  eprint = {2412.14858},
  archivePrefix = {arXiv},
  url = {https://arxiv.org/abs/2412.14858}
}

@article{ZerosExpansions,
  author = {H. Huang and Y. Yang and K. Zhou},
  title = {{Note on hidden zeros and expansions of tree-level amplitudes}},
  journal = {Eur. Phys. J. C},
  volume = {85},
  pages = {685},
  year = {2025},
  eprint = {2502.07173},
  archivePrefix = {arXiv},
  url = {https://arxiv.org/abs/2502.07173}
}

@article{HigherDerivativeZeros,
  author = {K. Zhou},
  title = {{Hidden zeros for higher-derivative YM and GR amplitudes at tree-level}},
  journal = {J. High Energy Phys.},
  volume = {2026},
  number = {02},
  year = {2026},
  pages = {039},
  eprint = {2510.11070},
  archivePrefix = {arXiv},
  url = {https://arxiv.org/abs/2510.11070}
}

@article{MassiveZeros,
  author = {Carrillo Gonz{\'a}lez, M. and F. Ward},
  title = {{Hidden zeros in massive field theories}},
  journal = {Phys. Rev. D},
  volume = {113},
  pages = {125018},
  year = {2026},
  eprint = {2601.16860},
  archivePrefix = {arXiv},
  url = {https://arxiv.org/abs/2601.16860}
}

@misc{FermionicZeros,
  author = {Y. Zhang and P. Zhu},
  title = {{Fermionic hidden zeros}},
  year = {2026},
  eprint = {2609.06552},
  archivePrefix = {arXiv},
  url = {https://arxiv.org/abs/2609.06552}
}

@misc{StringAllZeros,
  author = {Y.-C. Chang and H.-Y. Chen and Y.-t. Huang},
  title = {{All zeros of (super)String Theory}},
  year = {2025},
  eprint = {2506.15299},
  archivePrefix = {arXiv},
  url = {https://arxiv.org/abs/2506.15299}
}

@misc{IntegerRectangles,
  author = {S. He and Y. Zhang and F. Zhu and P. Zhu},
  title = {{String Amplitudes on Integer Rectangles}},
  year = {2026},
  eprint = {2610.31197},
  archivePrefix = {arXiv},
  url = {https://arxiv.org/abs/2610.31197}
}

@article{StringKLTUniversality,
  author = {C. Bartsch and K. Kampf and J. Novotn\'y and J. Trnka},
  title = {{Universality of colored scalars from the stringy Kawai--Lewellen--Tye kernel}},
  journal = {Phys. Rev. D},
  volume = {112},
  pages = {085002},
  year = {2025},
  eprint = {2505.01501},
  archivePrefix = {arXiv},
  url = {https://arxiv.org/abs/2505.01501}
}

@article{CosmologicalZeros,
  author = {S. De and S. Paranjape and A. Pokraka and M. Spradlin and A. Volovich},
  title = {{Hidden zeros of the cosmological wavefunction}},
  journal = {J. High Energy Phys.},
  volume = {2025},
  number = {07},
  year = {2025},
  pages = {174},
  eprint = {2503.23579},
  archivePrefix = {arXiv},
  url = {https://arxiv.org/abs/2503.23579}
}

@misc{LiRodinaCosmo,
  author = {Y. Li and L. Rodina},
  title = {{Cosmological Wavefunctions as Amplitudes: Dual Shuffle Factorization and Uniqueness from New Hidden Zeros}},
  year = {2026},
  eprint = {2604.01133},
  archivePrefix = {arXiv},
  url = {https://arxiv.org/abs/2604.01133}
}

@article{FormFactorsString,
  author = {Q. Cao},
  title = {{Form Factors from String Amplitudes}},
  journal = {Phys. Rev. Lett.},
  volume = {135},
  pages = {021603},
  year = {2025},
  eprint = {2504.15702},
  archivePrefix = {arXiv},
  url = {https://arxiv.org/abs/2504.15702}
}

@article{FormFactorSplitting,
  author = {L. Zhang},
  title = {{$2$-split of form factors via BCFW recursion relation}},
  journal = {J. High Energy Phys.},
  volume = {2026},
  number = {01},
  year = {2026},
  pages = {103},
  eprint = {2509.12564},
  archivePrefix = {arXiv},
  url = {https://arxiv.org/abs/2509.12564}
}

@article{CEGMSplitting,
  author = {Gim{\'e}nez Umbert, B. and B. Sturmfels},
  title = {{Splitting CEGM amplitudes}},
  journal = {J. High Energy Phys.},
  volume = {2025},
  number = {04},
  year = {2025},
  pages = {076},
  eprint = {2501.10805},
  archivePrefix = {arXiv},
  url = {https://arxiv.org/abs/2501.10805}
}

@article{CEGMNLSM,
  author = {N. Early},
  title = {{The CEGM NLSM}},
  journal = {J. High Energy Phys.},
  volume = {2025},
  number = {04},
  year = {2025},
  pages = {030},
  eprint = {2502.08016},
  archivePrefix = {arXiv},
  url = {https://arxiv.org/abs/2502.08016}
}

@article{NLSMZeroRecursion,
  author = {X. Li and K. Zhou},
  title = {{A new recursion relation for tree-level NLSM amplitudes based on hidden zeros}},
  journal = {J. High Energy Phys.},
  volume = {2026},
  number = {01},
  year = {2026},
  pages = {010},
  eprint = {2508.12894},
  archivePrefix = {arXiv},
  url = {https://arxiv.org/abs/2508.12894}
}

@article{OpenStringRepresentations,
  author = {N. Arkani-Hamed and C. Figueiredo and G. N. Remmen},
  title = {{Open string amplitudes: singularities, asymptotics and new representations}},
  journal = {J. High Energy Phys.},
  volume = {2025},
  number = {04},
  year = {2025},
  pages = {039},
  eprint = {2412.20639},
  archivePrefix = {arXiv},
  url = {https://arxiv.org/abs/2412.20639}
}

@article{RodinaTree,
  author = {L. Rodina},
  title = {{Hidden Zeros Are Equivalent to Enhanced Ultraviolet Scaling and Lead to Unique Amplitudes in $\mathrm{Tr}(\phi^3)$ Theory}},
  journal = {Phys. Rev. Lett.},
  volume = {134},
  pages = {031601},
  year = {2025},
  eprint = {2406.04234},
  archivePrefix = {arXiv},
  url = {https://arxiv.org/abs/2406.04234}
}

@article{BR,
  author = {J. V. Backus and L. Rodina},
  title = {{Emergence of Unitarity and Locality from Hidden Zeros at One-Loop Order}},
  journal = {Phys. Rev. Lett.},
  volume = {135},
  pages = {131601},
  year = {2025},
  eprint = {2503.03805},
  archivePrefix = {arXiv},
  url = {https://arxiv.org/abs/2503.03805}
}

@misc{CombinatorialLocality,
  author = {L. Rodina and G. Sellaroli and Y. Ni},
  title = {{A Combinatorial Origin of Locality}},
  year = {2026},
  eprint = {2608.23008},
  archivePrefix = {arXiv},
  url = {https://arxiv.org/abs/2608.23008}
}

@article{ART,
  author = {N. Arkani-Hamed and L. Rodina and J. Trnka},
  title = {{Locality and Unitarity of Scattering Amplitudes from Singularities and Gauge Invariance}},
  journal = {Phys. Rev. Lett.},
  volume = {120},
  pages = {231602},
  year = {2018},
  eprint = {1612.02797},
  archivePrefix = {arXiv},
  url = {https://arxiv.org/abs/1612.02797}
}

@article{RodinaAdler,
  author = {L. Rodina},
  title = {{Uniqueness from gauge invariance and the Adler zero}},
  journal = {J. High Energy Phys.},
  volume = {2019},
  number = {09},
  year = {2019},
  pages = {084},
  eprint = {1612.06342},
  archivePrefix = {arXiv},
  url = {https://arxiv.org/abs/1612.06342}
}

@article{EFTSoft,
  author = {C. Cheung and K. Kampf and J. Novotny and J. Trnka},
  title = {{Effective Field Theories from Soft Limits of Scattering Amplitudes}},
  journal = {Phys. Rev. Lett.},
  volume = {114},
  pages = {221602},
  year = {2015},
  eprint = {1412.4095},
  archivePrefix = {arXiv},
  url = {https://arxiv.org/abs/1412.4095}
}

@article{RodinaSoft,
  author = {L. Rodina},
  title = {{Scattering Amplitudes from Soft Theorems and Infrared Behavior}},
  journal = {Phys. Rev. Lett.},
  volume = {122},
  pages = {071601},
  year = {2019},
  eprint = {1807.09738},
  archivePrefix = {arXiv},
  url = {https://arxiv.org/abs/1807.09738}
}

@article{RodinaBCFW,
  author = {L. Rodina},
  title = {{Uniqueness from locality and BCFW shifts}},
  journal = {J. High Energy Phys.},
  volume = {2019},
  number = {09},
  year = {2019},
  pages = {078},
  eprint = {1612.03885},
  archivePrefix = {arXiv},
  url = {https://arxiv.org/abs/1612.03885}
}

@article{CarrascoRodinaUV,
  author = {J. J. M. Carrasco and L. Rodina},
  title = {{UV considerations on scattering amplitudes in a web of theories}},
  journal = {Phys. Rev. D},
  volume = {100},
  pages = {125007},
  year = {2019},
  eprint = {1908.08033},
  archivePrefix = {arXiv},
  url = {https://arxiv.org/abs/1908.08033}
}

@article{RodinaCHYUV,
  author = {L. Rodina},
  title = {{UV consistency conditions for Cachazo-He-Yuan integrands}},
  journal = {Phys. Rev. D},
  volume = {102},
  pages = {045012},
  year = {2020},
  eprint = {2005.06446},
  archivePrefix = {arXiv},
  url = {https://arxiv.org/abs/2005.06446}
}

@article{Kontsevich,
  author = {M. Kontsevich},
  title = {{Intersection theory on the moduli space of curves and the matrix Airy function}},
  journal = {Commun. Math. Phys.},
  volume = {147},
  pages = {1--23},
  year = {1992},
  doi = {10.1007/BF02099526},
  url = {https://doi.org/10.1007/BF02099526}
}

@misc{SM,
  note = {See Appendices~\ref{app:boundary}--\ref{app:four} for the zero assignments and reconstruction arguments. The accompanying Wolfram Language package \texttt{AiryUniqueness.wl} checks $L=2,3$ at $n=1,2,3$. The four-loop assignments and finite checks are described in Appendix~\ref{app:four}.}
}

@article{Counting,
  author = {N. Arkani-Hamed and H. Frost and G. Salvatori and P.-G. Plamondon and H. Thomas},
  title = {{All loop scattering as a counting problem}},
  journal = {J. High Energy Phys.},
  volume = {2025},
  number = {08},
  year = {2025},
  pages = {194},
  eprint = {2309.15913},
  archivePrefix = {arXiv},
  url = {https://arxiv.org/abs/2309.15913}
}

@article{AllMultiplicity,
  author = {N. Arkani-Hamed and H. Frost and G. Salvatori and P.-G. Plamondon and H. Thomas},
  title = {{All loop scattering for all multiplicity}},
  journal = {J. High Energy Phys.},
  volume = {2025},
  number = {09},
  year = {2025},
  pages = {033},
  eprint = {2311.09284},
  archivePrefix = {arXiv},
  url = {https://arxiv.org/abs/2311.09284}
}

@article{ScalarScaffold,
  author = {N. Arkani-Hamed and Q. Cao and J. Dong and C. Figueiredo and S. He},
  title = {{Scalar-scaffolded gluons and the combinatorial origins of Yang--Mills theory}},
  journal = {J. High Energy Phys.},
  volume = {2025},
  number = {04},
  year = {2025},
  pages = {078},
  eprint = {2401.00041},
  archivePrefix = {arXiv},
  url = {https://arxiv.org/abs/2401.00041}
}

@article{TropicalLagrangians,
  author = {N. Arkani-Hamed and C. Figueiredo and H. Frost and G. Salvatori},
  title = {{Tropical amplitudes for colored Lagrangians}},
  journal = {J. High Energy Phys.},
  volume = {2025},
  number = {05},
  year = {2025},
  pages = {051},
  eprint = {2402.06719},
  archivePrefix = {arXiv},
  url = {https://arxiv.org/abs/2402.06719}
}

@article{Surface,
  author = {N. Arkani-Hamed and Q. Cao and J. Dong and C. Figueiredo and S. He},
  title = {{Surface Kinematics and the Canonical Yang-Mills All-Loop Integrand}},
  journal = {Phys. Rev. Lett.},
  volume = {134},
  pages = {171601},
  year = {2025},
  eprint = {2408.11891},
  archivePrefix = {arXiv},
  url = {https://arxiv.org/abs/2408.11891}
}

@article{InverseKLTSurface,
  author = {C. Bartsch and K. Kampf and J. Novotn\'y and J. Trnka},
  title = {{A surface integrand for the inverse KLT kernel}},
  journal = {J. High Energy Phys.},
  volume = {2026},
  number = {10},
  year = {2026},
  pages = {099},
  eprint = {2602.15102},
  archivePrefix = {arXiv},
  url = {https://arxiv.org/abs/2602.15102}
}

@article{NLSMShift,
  author = {N. Arkani-Hamed and Q. Cao and J. Dong and C. Figueiredo and S. He},
  title = {{Nonlinear sigma model amplitudes to all loop orders are contained in the $\mathrm{Tr}(\Phi^3)$ theory}},
  journal = {Phys. Rev. D},
  volume = {110},
  pages = {065018},
  year = {2024},
  eprint = {2401.05483},
  archivePrefix = {arXiv},
  url = {https://arxiv.org/abs/2401.05483}
}

@article{CirclesTriangles,
  author = {N. Arkani-Hamed and C. Figueiredo},
  title = {{Circles and triangles, the NLSM and $\mathrm{Tr}(\Phi^3)$}},
  journal = {J. High Energy Phys.},
  volume = {2025},
  number = {09},
  year = {2025},
  pages = {189},
  eprint = {2403.04826},
  archivePrefix = {arXiv},
  url = {https://arxiv.org/abs/2403.04826}
}

@article{ZhouDiagrams,
  author = {K. Zhou},
  title = {{Understanding zeros and splittings of ordered tree amplitudes via Feynman diagrams}},
  journal = {J. High Energy Phys.},
  volume = {2025},
  number = {03},
  year = {2025},
  pages = {154},
  eprint = {2411.07944},
  archivePrefix = {arXiv},
  url = {https://arxiv.org/abs/2411.07944}
}

@article{UniversalDiagrammaticZeros,
  author = {K. Zhou},
  title = {{Universal interpretation of hidden zero and $2$-split of tree-level amplitudes using Feynman diagrams. Part I. $\mathrm{Tr}(\phi^3)$, NLSM and YM}},
  journal = {J. High Energy Phys.},
  volume = {2026},
  number = {09},
  year = {2026},
  pages = {112},
  eprint = {2604.23680},
  archivePrefix = {arXiv},
  url = {https://arxiv.org/abs/2604.23680}
}

@article{SplittingDiagramsExpansions,
  author = {B. Feng and L. Zhang and K. Zhou},
  title = {{2-split from Feynman diagrams and expansions}},
  journal = {J. High Energy Phys.},
  volume = {2026},
  number = {02},
  year = {2026},
  pages = {204},
  eprint = {2508.21345},
  archivePrefix = {arXiv},
  url = {https://arxiv.org/abs/2508.21345}
}

@article{SplittingBCFW,
  author = {B. Feng and L. Zhang and K. Zhou},
  title = {{Hidden zeros and 2-split via BCFW recursion relation}},
  journal = {J. High Energy Phys.},
  volume = {2025},
  number = {08},
  year = {2025},
  pages = {205},
  eprint = {2504.14215},
  archivePrefix = {arXiv},
  url = {https://arxiv.org/abs/2504.14215}
}

@article{ZhouLoopZeros,
  author = {K. Zhou},
  title = {{Towards New Hidden Zero and $2$-Split of Loop-Level Feynman Integrands in $\mathrm{Tr}(\phi^3)$ Model}},
  journal = {J. High Energy Phys.},
  volume = {2026},
  number = {10},
  year = {2026},
  pages = {030},
  eprint = {2604.13810},
  archivePrefix = {arXiv},
  url = {https://arxiv.org/abs/2604.13810}
}

@article{WittenGravity,
  author = {E. Witten},
  title = {{Two-dimensional gravity and intersection theory on moduli space}},
  journal = {Surveys Diff. Geom.},
  volume = {1},
  pages = {243--310},
  year = {1991},
  doi = {10.4310/SDG.1990.v1.n1.a5},
  url = {https://doi.org/10.4310/SDG.1990.v1.n1.a5}
}

@article{GrossNewman,
  author = {D. J. Gross and M. J. Newman},
  title = {{Unitary and Hermitian matrices in an external field. II: The Kontsevich model and continuum Virasoro constraints}},
  journal = {Nucl. Phys. B},
  volume = {380},
  pages = {168--180},
  year = {1992},
  eprint = {hep-th/9112069},
  archivePrefix = {arXiv},
  url = {https://arxiv.org/abs/hep-th/9112069}
}

@article{PolynomialAverages,
  author = {Di Francesco, P. and C. Itzykson and J.-B. Zuber},
  title = {{Polynomial averages in the Kontsevich model}},
  journal = {Commun. Math. Phys.},
  volume = {151},
  pages = {193--219},
  year = {1993},
  eprint = {hep-th/9206090},
  archivePrefix = {arXiv},
  url = {https://arxiv.org/abs/hep-th/9206090}
}

\onecolumngrid
\clearpage
\appendix
\setcounter{figure}{0}
\renewcommand{\thefigure}{A\arabic{figure}}
\setcounter{table}{0}
\renewcommand{\thetable}{A\arabic{table}}
\renewcommand{\theHfigure}{app.\arabic{figure}}
\renewcommand{\theHtable}{app.\arabic{table}}
\twocolumngrid

\section{Boundary zeros, core reduction and lifting}
\label{app:boundary}\label{app:reflection}
We give the boundary argument at arbitrary multiplicity. Summing all insertions of one external leg proves the reference zero; comparing pole coefficients of the weighted sum then relates its parent diagrams. We use these relations to reduce the ansatz to core weights and to lift the finite uniqueness results.

\begin{figure*}[!t]
 \centering
 \begin{minipage}[t]{.68\textwidth}
 \vspace{0pt}\centering
 \includegraphics[width=\linewidth]{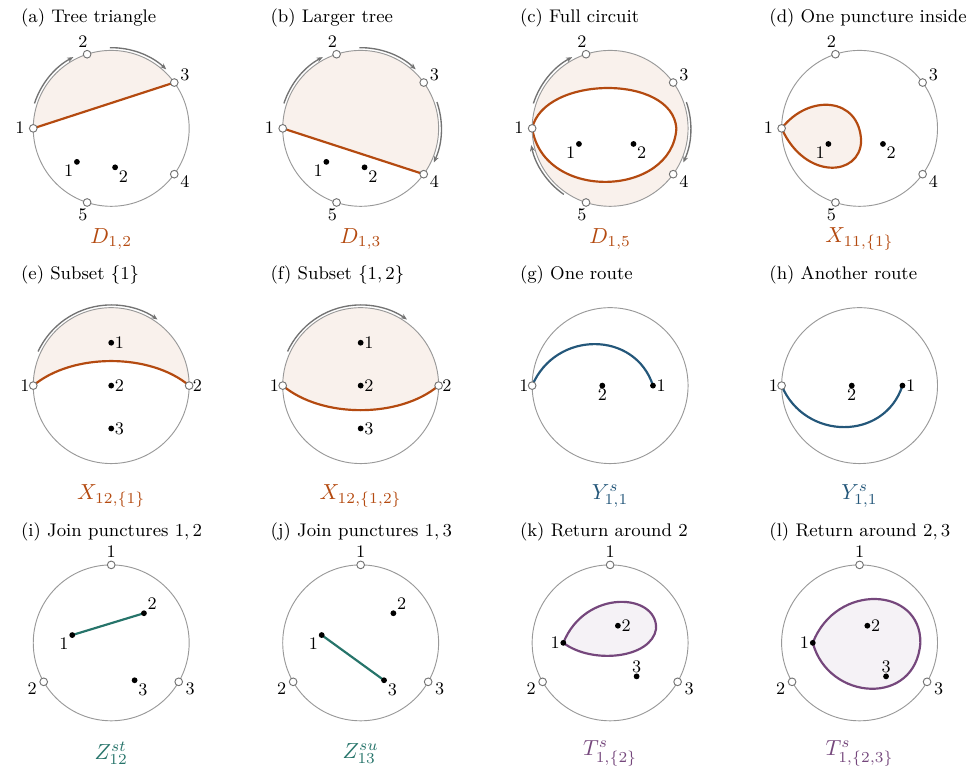}
 \caption{The channel variables. Open circles are boundary marks; filled dots are punctures. Shading selects the boundary-interval side of $D,X$ and the enclosed subset of $T$. Thus (c) shades the outside of the full-circuit $D_{1,5}$, while (d) shades the inside of $X_{11,\{1\}}$. The winding routes in (g,h) share $Y_{1,1}^s$.}
 \label{fig:projection-examples}\label{fig:channel-dictionary}
 \end{minipage}\hfill
 \begin{minipage}[t]{.28\textwidth}
 \vspace{0pt}\centering
 \includegraphics[width=\linewidth]{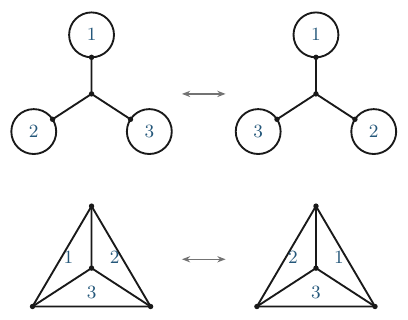}
 \caption{The two three-loop mirror pairs. Each face retains its label and sign. Reflection reverses the boundary order while transporting the curves and their chosen sides. The two pairs reduce 61 core weights to 59 per sign sector.}
 \label{fig:reflection-example}
 \end{minipage}
\end{figure*}

\subsection{The boundary identity at arbitrary multiplicity}

We prove that the reference integrand satisfies the boundary zero
at any multiplicity and loop order. The proof groups diagrams
according to the graph obtained by removing one external leg.
Within each group, the diagrams differ only in where that leg
is attached, and their propagator products cancel by telescoping.
We work throughout in a fixed puncture-sign sector.

\paragraph{The zero locus.}
For the boundary zero associated with adjacent marks $n-1,n$,
choose $\eta\ne0$. At $n\ge3$, impose the following relations
among the $n$-point inverse propagators:
\begin{equation}
 \begin{aligned}
 Y_{n,p}^s-Y_{n-1,p}^s&=\eta,\\
 X_{i,n,S}-X_{i,n-1,S}&=\eta
 \quad (1\le i\le n-2),\\
 X_{n-1,n,S}-X_{n-1,n-1,S}&=\eta,\\
 X_{n,n,S}-X_{n-1,n-1,S}&=2\eta,\\[2pt]
 D_{n-1,2}&=-\eta,\\ D_{n-2,2}&=\eta,\\
 D_{i,n-i}-D_{i,n-1-i}&=\eta
 \quad (1\le i\le n-3),\\
 D_{n,d}-D_{n-1,d+1}&=\eta
\quad  (2\le d\le n-1),\\
 D_{n,n}-D_{n-1,n}&=2\eta.
 \end{aligned}
 \label{eq:boundary}
\end{equation}
The $Y,X$ relations apply to every puncture, puncture sign
and allowed subset $S$. All remaining channels, including
every $Z,T$, are free. At two points, use the same $Y,X$
relations and set $D_{1,2}=-\eta$, $D_{2,2}=\eta$.
Cyclic relabelling gives the other boundary zeros.
All propagators are kept nonzero.

\paragraph{Grouping the diagrams.}
Remove the selected external leg from an $n$-point diagram. The result is an $(n-1)$-point graph at the same
loop order,   which we call the child. All diagrams that give
this same child form one group of parents. Hatted variables
used below denote channels of the child.

\begin{figure}[!htb]
 \centering
 \includegraphics[width=.94\linewidth]{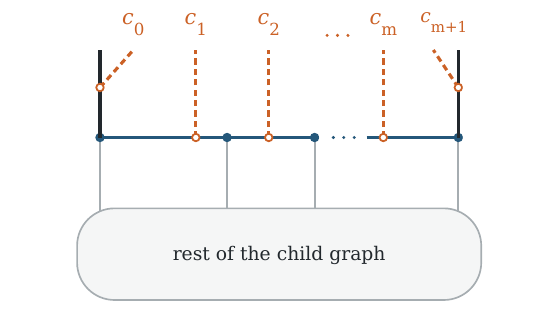}
 \caption{All allowed positions of one new external leg
 between two fixed neighboring external particles.
 Each orange segment represents a separate parent diagram.
 The interior positions split an existing propagator.
 The two end positions attach to a neighboring external leg
 and create a two-particle tree. Both end insertions point
 inward to preserve the cyclic ordering. The rest of the
 child graph is arbitrary.}
 \label{fig:boundary-ordered-insertions}
\end{figure}

To recover the parents, insert the removed leg in every
position compatible with its original place in the cyclic
external ordering. The new leg must remain between the
same two neighboring external particles. In the child,
these particles bound a segment of the exterior face,
shown schematically in Fig.~\ref{fig:boundary-ordered-insertions}.

There are two kinds of insertion. On an internal edge along
this segment, add a cubic vertex and attach the new external
leg to it. The original propagator is then replaced by two
propagators. Alternatively, attach the new leg to either
of the two neighboring external legs. This creates a cubic
vertex carrying two external legs, connected to the rest
of the graph by a new propagator. These are the two
external-tree attachments, with channels
$D_{n-1,2}=-\eta$ and $D_{n-2,2}=\eta$.

Number the internal-edge insertions $1,\ldots,m$, starting
from the end with tree channel $-\eta$. Label the two
external-tree insertions $0$ and $m+1$. If an edge occurs
twice along this boundary, its two insertion positions
are counted separately. For a one-point child, the two
end positions are on opposite sides of its sole external leg.

\paragraph{The telescoping cancellation.}
Let $P_0$ be the product of propagators in the first parent
after removing its new tree factor $1/D_{n-1,2}$.
Now pass through the ordered internal insertion positions.
At each position, the relations in Eq.~\eqref{eq:boundary}
change one inverse propagator by $\eta$. Define $P_k$
from $P_{k-1}$ by making this change at position $k$.

More explicitly, let $Q_k$ be the inverse propagator before
this change, and let $B_k$ be the product of all the other
propagators. Then
\[
 P_{k-1}=\frac{B_k}{Q_k},
 \qquad
 P_k=\frac{B_k}{Q_k+\eta}.
\]
The parent obtained by inserting the leg at position $k$
contains both propagators $1/Q_k$ and $1/(Q_k+\eta)$.
Its contribution is therefore
\begin{equation}
 A_k=\frac{B_k}{Q_k(Q_k+\eta)}
     =\frac{P_{k-1}-P_k}{\eta}.
 \label{eq:app-insertion-difference}
\end{equation}
Thus each internal insertion gives the difference of two
successive products. The negative $P_k$ from position $k$
cancels the positive $P_k$ from position $k+1$.

The two external-tree parents supply the terms needed at
the ends of this sequence:
\[
 A_0=-\frac{P_0}{\eta},
 \qquad
 A_{m+1}=\frac{P_m}{\eta}.
\]
Consequently the whole group vanishes,
\begin{equation}
 \sum_{k=0}^{m+1}A_k
 =-\frac{P_0}{\eta}
  +\sum_{k=1}^{m}\frac{P_{k-1}-P_k}{\eta}
  +\frac{P_m}{\eta}
 =0.
 \label{eq:app-boundary-cancellation}
\end{equation}

The child graph was arbitrary, so the cancellation holds
independently of its multiplicity and loop structure.
Every parent belongs to the group selected by deleting
the chosen leg. Summing these vanishing groups therefore
proves the boundary zero for the full reference integrand.
When different diagrams have the same projected denominator
product, their contributions combine with their graph
multiplicities.

For the lifting argument, we also need the same sum with
$D_{n-1,2}=s_0$ left independent. Keep every other relation
in Eq.~\eqref{eq:boundary} unchanged. Only the first
external-tree contribution changes, giving
\begin{equation}
 \frac{P_0}{s_0}
 +\sum_{k=1}^{m}\frac{P_{k-1}-P_k}{\eta}
 +\frac{P_m}{\eta}
 =\left(\frac1{s_0}+\frac1\eta\right)P_0.
 \label{eq:app-telescope}
\end{equation}
Summing over children and puncture signs gives
\begin{equation}
 \left.\mathcal I_{n,L}\right|_{\rm split}
 =\left(\frac1{s_0}+\frac1\eta\right)
 \mathcal I_{n-1,L}(\widehat Q),
 \label{eq:app-tree-loop-split}
\end{equation}
where $\widehat Q$ are the child channel values. This is the
four-point-tree specialization of the tree--loop split in
Ref.~\cite{Splits}. The prefactor is the cubic tree amplitude
$\mathcal A_4(s_0,\eta)=1/s_0+1/\eta$. Setting $s_0=-\eta$
therefore gives every boundary zero used here, at arbitrary
multiplicity and loop order, with the lower-point factor regular.
Keeping $s_0$ independent gives the insertion identity used
in the lifting argument below.

\subsection{Coefficient comparisons for arbitrary parents}
We now apply the identity to unknown diagram coefficients. Take residues in unchanged propagators to select the child structure being compared. Weight its interior insertions by $c_1,\ldots,c_m$ and its two external-tree insertions by $c_0,c_{m+1}$. Equation~\eqref{eq:app-telescope} at $s_0=-\eta$ becomes
\begin{equation}
 \frac1\eta\left[(c_1-c_0)P_0+
 \sum_{k=1}^{m-1}(c_{k+1}-c_k)P_k
 +(c_{m+1}-c_m)P_m\right]=0.
 \label{eq:app-weighted-path}
\end{equation}
Distinct shifted poles compare neighboring coefficients. When the products are distinguished by these poles, the comparisons give $c_0=c_1=\cdots=c_{m+1}$. This is the tree-level diagram-subset argument~\cite{RodinaTree} applied to an arbitrary loop-containing child.

The winding projection can identify some of the products in Eq.~\eqref{eq:app-weighted-path}. We then use another deleted leg or another pole to separate the coefficients. At low multiplicity there may be too few such comparisons. The next step determines when the boundary equations nevertheless fix every attachment to a given core.

\subsection{Reducing the remaining freedom to core weights}
\label{app:finite-reduction}\label{app:remainder}

We now show that the boundary equations leave only the
weights of complete core attachment sums, once the
multiplicity is sufficiently large. We first determine
the terms containing external trees, then exclude any
remaining terms.

\paragraph{Terms containing external trees.}
A residue in $D_{i,2}$ removes a two-particle tree and
replaces it by one external leg. On a boundary zero, two
neighboring triangle channels take the values $\eta$
and $-\eta$. They cannot occur in the same triangulation,
so the coefficient of $1/\eta$ equates their residues,
with the remaining particles labelled in the same order.
Repeating this comparison relates all triangular residues.

These residues satisfy the lower-point boundary equations.
For example, for $n\ge4$ the boundary assignment gives
\begin{equation}
 D_{n,2}=Q+\eta,\qquad D_{n-1,3}=Q.
 \label{eq:app-boundary-descent}
\end{equation}
Here $Q$ is an independent channel value. Taking the pole
at $Q=-\eta$ selects the residue in $D_{n,2}$. After removing
that two-particle tree, the remaining propagators satisfy
the boundary assignment at $n-1$ points. This also applies
at four points.

Suppose, by induction, that the common residue is a linear
combination of complete core sums. Attach the removed
particle in all positions on those same cores, with the
same weights, and subtract the resulting $n$-point sum
from the ansatz. The difference has zero residue in every
$D_{i,2}$. Restoring a specified two-particle tree is unique,
so these residues eliminate every term containing a triangle
channel. Every external tree contains two external legs
meeting at a cubic vertex, and hence such a channel.
The difference therefore contains no $D$ propagators.

\paragraph{Terms without external trees.}
Consider a term with three consecutive legs attached to
the same side of a core edge. Delete the middle leg and
apply the insertion comparisons of
Eq.~\eqref{eq:app-weighted-path}. The retained neighbors
fix the insertion order. Since the external-tree terms
are absent, the two end coefficients are
$c_0=c_{m+1}=0$, and the pole equations force the intervening
coefficients to vanish as well. When propagators repeat,
the same comparison uses the highest-order poles after
partial fractions in the lower-point channels.

A surviving term can thus have at most two legs on each
side of a core edge facing the exterior. A cubic vacuum
core has $3L-3$ edges, with $6L-6$ sides in total. Its $L$
internal faces use at least one side each, leaving at
most $5L-6$ exterior sides. Consequently, for $L\ge2$,
a nonzero remainder is possible only when
\begin{equation}
 n\le N_L:=2(5L-6)=10L-12.
 \label{eq:general-finite-bound}
\end{equation}

It remains to solve a finite set of boundary equations.
For each core, distribute at most two legs on each
exterior side and substitute every cyclic placement of
Eq.~\eqref{eq:boundary}. Partial fractions in the independent
channel variables give a linear equation for each distinct
pole product. Full column rank of this system means that
no nonzero combination of the remaining diagrams survives.
The actual starting multiplicity $n_L$ is the point
above which these equations exclude every remainder;
it can be much smaller than $N_L$.

\paragraph{The starting multiplicities.}
At two loops, the finite equations exclude all remainders
for $3\le n\le8$.

At three loops, the five topologies in
Fig.~\ref{fig:app-three-cores} permit a direct case analysis.
First consider cores without a bridge bordering the
exterior face. If only one puncture borders that face,
the boundary assignment leaves its rays arbitrary, so
the remainder vanishes. If two punctures do, choose $q$
to occur on only one exterior side and call the other
$p$. After clearing denominators, the numerator is at
most linear in each $Y_{i,q}$. On each cyclic boundary
zero,
\[
 d_i=d_{i+1},\qquad d_i=Y_{i,q}-Y_{i,p}.
\]
The numerator must therefore contain
$\prod_i(d_i-d_{i+1})$, which is quadratic in each
$Y_{i,q}$ for $n\ge3$. It must be zero.

For the exterior triangle, at most two legs per side
leaves only $(2,2,0)$, $(2,1,1)$, $(2,2,1)$ and $(2,2,2)$
above three points. Deleting an end leg and comparing
pole coefficients eliminates these placements.

The four sides of the remaining core border internal faces
$1,2,3,2$, in order around the exterior. The same comparisons
remove all placements except one with a leg on each side.
For this last case, apply the boundary zero at marks $3,4$.
After suppressing the common
internal factors and the rays at marks $1,2$, the two
products contributing to the relevant pole are
\begin{equation}
 \frac{c_A}{Y_{3,2}Y_{3,3}Y_{4,2}Y_{4,3}}
 +\frac{c_B}{(Y_{3,2})^2Y_{3,3}Y_{4,2}}.
 \label{eq:app-four-side-comparison}
\end{equation}
Set $Y_{4,2}=Y_{3,2}+\eta$ and $Y_{4,3}=Y_{3,3}+\eta$.
The coefficient of $1/(Y_{3,2}Y_{3,3})$ in partial fractions
is $(c_A-c_B)/\eta^2$, so $c_A=c_B$. The second product has
two legs on one side bordering puncture $2$ and was already
eliminated by the preceding pole comparisons. Hence the
last coefficient vanishes as well.

For cores with bridges, the same insertion argument
limits each attached one-puncture component to two legs.
In a chain, the pole comparisons can fail only when the
deleted leg was the sole attachment to an end component.
There are at most two such legs. For $n\ge4$, another
choice of deleted leg therefore sets the coefficient
to zero.

At four loops, the bound of two legs applies to an entire
tadpole together with the edge connecting it to the rest
of the core. This branch occupies three exterior sides:
the tadpole and the two sides of its connecting edge.
If a core has $m$ exterior sides and $t$ such disjoint
branches, the bound is therefore
\begin{equation}
 n\le2(m-3t)+2t=2(m-2t).
 \label{eq:app-four-attachment-bound}
\end{equation}
The four-loop core list gives $m-2t\le8$, reducing the
range from $N_4=28$ to $n\le16$. Applying the finite
boundary equations above to these allowed distributions
gives full column rank modulo 65521 for $4\le n\le16$,
and hence also over the rationals. These remainder
calculations are separate from the low-point uniqueness
checks in the accompanying WL package.

Together with the finite starting checks, these arguments
give $n_2=n_3=3$ and $n_4=4$. At one loop the remainder
is simply $c/\prod_iY_{i,1}^s$, which a boundary zero
forces to vanish, giving $n_1=2$.

Below these multiplicities, boundary zeros can leave
additional coefficients. At two loops the boundary
equations leave $7,5,4$ weights per sign at $n=1,2,3$,
while the core sums have dimension four. At four loops,
the three-point boundary space has $1644=1624+20$ weights
per sign, whereas four points leave precisely the
1624 core weights.

\subsection{Passing a loop zero to the residue}
We now explain how to construct the higher-point loop conditions used in the main-text induction. The boundary equations have already restricted the parent ansatz. Apply the insertion assignment to $M_{n+1,L}$, keeping the new triangle channel $s_0$ independent. Only terms containing this triangle depend on $s_0$, and it occurs once in each such term. Hence the result is
\begin{equation}
 \left.M_{n+1,L}\right|_{\rm split}
 =\frac{R_{n,L}(\widehat Q)}{s_0}+C,
 \label{eq:app-residue-plus-rest}
\end{equation}
where $R_{n,L}$ is the triangle residue and $C$ does not depend on $s_0$. The notation means substitution of Eq.~\eqref{eq:boundary}, with $n+1$ parent particles. Its boundary-zero value $s_0=-\eta$ fixes $C=R_{n,L}/\eta$. Therefore
\begin{equation}
 \left.M_{n+1,L}\right|_{\rm split}
 =\left(\frac1{s_0}+\frac1\eta\right)R_{n,L}(\widehat Q).
 \label{eq:unknown-released-split}
\end{equation}
This uses the boundary equations to express the other insertions in terms of the residue.

Choose any required $n$-point loop zero and put the hatted channels on it. For the reference sum, $R_{n,L}=\mathcal I_{n,L}$, so the right-hand side vanishes. This constructs a valid $(n+1)$-point zero. Impose it on the unknown parent with $s_0\ne-\eta$. The prefactor is nonzero, and the residue must obey the chosen lower-point loop zero. Repeating for all required families puts the residue under the full lower-point uniqueness conditions.

Finally, boundary reduction has written the parent as $\sum_r w_{r,n+1}C_{r,n+1,L}$. Removing a two-particle tree from a complete attachment sum gives the lower-point sum on the same core, with the same parent weight. Thus lower-point uniqueness gives
\begin{equation}
 \sum_r w_{r,n+1}C_{r,n,L}
 =a_{n+1,L}\mathcal I_{n,L}.
 \label{eq:app-core-weight-descent}
\end{equation}
For $n\ge n_L$ these core sums are independent, so each parent weight is fixed to its reference value times $a_{n+1,L}$. This fixes all parent attachments, including those without a $D$ propagator. The argument concerns the weights of this parent ansatz; its overall normalization is still free at each multiplicity.

\section{Explicit loop-zero families}
\label{app:families}
We first explain how external insertions extend the Airy zero to arbitrary multiplicity, then list the larger families containing the main-text zeros and the additional assignments used at low multiplicity. The construction starts from the Airy face sums. Connections between punctures and the returns around them are varied together; cutting a return identifies groups whose cancellation allows that propagator to remain free. The prescriptions below specify the resulting zeros. All free parameters are chosen away from propagator poles, and a sign exchange acts on every channel ending at the chosen puncture.

\subsection{Adding external legs to the Airy zero}
\label{app:airy-insertions}
We show how the vacuum Airy cancellation gives the all-$n$ assignment in Eq.~\eqref{eq:airy-locus-main}. We first attach legs directly to the loop graph, without external trees. Let $H_{n,L}$ be the sum of these diagrams on the Airy assignment in a fixed puncture-sign sector, and let $A_{L+1}$ be the corresponding vacuum sum. Their internal-face weights are left implicit.

Adding the first leg splits one outer edge into two. The factor $1/(b+sa_p)$ becomes $1/(b+sa_p)^2=-\partial_b[1/(b+sa_p)]$. Differentiating the whole vacuum sum therefore adds a leg at every position along the outer face. An edge bordering that face on both sides contributes twice, once for each side. Thus $H_{1,L}=-\partial_b A_{L+1}$.

\begin{figure}[!htb]
 \centering
 \includegraphics[width=.94\linewidth]{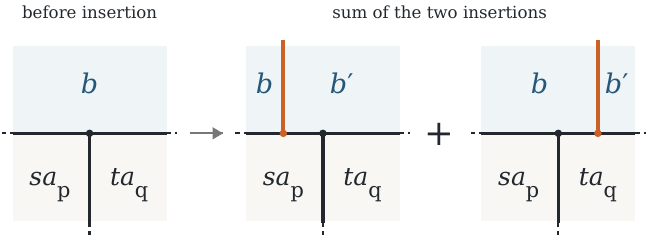}
 \caption{Adding one external leg to two adjacent edges on the outer face. The orange leg can attach to either edge, dividing the boundary region of weight $b$ into regions of weights $b,b'$. The internal faces retain weights $sa_p,ta_q$, and the rest of the graph is unchanged. Summing these two insertions gives the difference of the original two-edge product at $b$ and $b'$, divided by $b'-b$. The same cancellation applies to any number of adjacent edges.}
 \label{fig:airy-edge-insertion}
\end{figure}

Adding another leg divides one boundary region into two, to which we assign independent weights $b,b'$. The split edge now contributes
\begin{equation}
 \frac{1}{(b+sa_p)(b'+sa_p)}
 =\frac{1}{b'-b}\left(\frac{1}{b+sa_p}-\frac{1}{b'+sa_p}\right).
 \label{eq:airy-leg-insertion-main}
\end{equation}
This is why differences appear: they express the two propagators created by inserting a leg. Figure~\ref{fig:airy-edge-insertion} shows the two possible insertions on a pair of adjacent edges. Applying Eq.~\eqref{eq:airy-leg-insertion-main} to the split edge in each diagram cancels the mixed product $1/[(b+sa_p)(b'+ta_q)]$. Only the original product evaluated at $b$ and at $b'$ remains. The same cancellation continues along a longer sequence of edges, giving
\begin{equation}
 H_{n+1,L}(\ldots,b,b')=
 \frac{H_{n,L}(\ldots,b)-H_{n,L}(\ldots,b')}{b'-b}.
 \label{eq:airy-successive-insertions-main}
\end{equation}

Deleting the new leg reverses the construction, so the insertions generate every higher-point diagram with the chosen cyclic ordering. We repeat this operation whenever we add a particle, introducing one new boundary weight each time. Every new edge still carries the sum of its adjacent face weights, so the resulting diagrams are precisely on the higher-point Airy assignment. The differences act only on boundary weights and preserve oddness in each internal-face weight. The puncture-sign sum therefore vanishes after every insertion. Coincident boundary weights follow by taking $b'\to b$, when the difference becomes a derivative.

At two loops, Eq.~\eqref{eq:airy-two-loop-main} gives the explicit example
\begin{equation}
 \begin{aligned}
 H_{1,2}^{st}(b)&=\frac{st}{2b^2a_pa_q},\\
 H_{2,2}^{st}(b,b')&=\frac{st(b+b')}{2b^2(b')^2a_pa_q}.
 \end{aligned}
 \label{eq:airy-two-insertions-example}
\end{equation}
Both retain the same factor $st$, so summing the four sign sectors gives zero. Further insertions preserve this sign dependence in the same way.

Finally, fix an external-tree structure. Removing its trees leaves one of these vanishing sums at the surviving attachment marks. Restoring the trees multiplies that sum by their product of $D^{-1}$ factors. Summing over tree structures proves the full zero, with every $D$ independent.

\subsection{Two loops}
\label{app:two}
The full mountain contains the two simple deformations used in the main text. For ordered punctures $p,q$, choose independent $u_i,v_i,h,z,e$ and assign

\begin{equation}
 \begin{array}{c|cc}
 &+&-\\\hline
 Y_{i,p}&u_i&v_i+h\\
 Y_{i,q}&u_i+h&v_i
 \end{array},\qquad
 (Z_{pq}^{st})=\begin{pmatrix}z&-h\\h&e\end{pmatrix},
 \label{eq:app-two-general-rays}
\end{equation}

\begin{equation}
 \begin{aligned}
 X_{ij,\{p\}}&=u_i+v_j+h,\\
 X_{ij,\{q\}}&=u_j+v_i+h,\\
 T_{p,\{q\}}^+&=z-h,&T_{p,\{q\}}^-&=e+h,\\
 T_{q,\{p\}}^+&=z+h,&T_{q,\{p\}}^-&=e-h.
 \end{aligned}
 \label{eq:app-two-general-internal}
\end{equation}

The rows and columns of $Z$ carry $s,t=+,-$; all $D$ remain free. Set $a_p=(z-h)/2$, $a_q=(z+h)/2$, $b_i=(u_i+v_i+h)/2$, $\xi_i=(u_i-v_i-z)/2$ and $\epsilon=e+z$. Airy is $\xi_i=\epsilon=0$. The common-shift family has $\xi_i=\xi$, $\epsilon=0$; exchanging the $q$ sign gives the opposite-shift family.

The internal rule is particularly simple: each return is a row or column sum of the $Z$ matrix. Allowing independent $u_i,v_i$ then separates the two boundary orientations while preserving the cancellation. The four core contributions cancel for arbitrary ordered external insertions, and restoring external trees leaves every $D$ free.

The common- and opposite-shift families suffice at every multiplicity used in the finite checks. At $n=1$, their 28 signed coefficients reduce to one; at $n=2$, both boundary placements leave 20 coefficients, which the two families again reduce to one. At $n=3$, they fix the sixteen remaining core weights. No additional low-point zero is needed.

\subsection{Three loops}
\label{app:three}\label{app:three-low}
We give the choices used in the main text and the larger families needed for the one- and two-point checks. Write $E_{\boldsymbol\gamma}$ for the uniform internal shift in Eq.~\eqref{eq:three-simple-zero-main}, with one selected sign $\gamma_p=\pm1$ per puncture. Write $R_{p,S}^s$ for a single-return displacement in Eq.~\eqref{eq:three-return-releases-main}. The six sufficient three-point choices are
\begin{equation}
 \begin{gathered}
 E_{+++},\ E_{++-},\ E_{+-+},\\
 R_{1,\{3\}}^{-},\ R_{2,\{1\}}^{+},\ R_{3,\{2\}}^{+}.
 \end{gathered}
 \label{eq:three-six-family-list}
\end{equation}
No further images are required for this choice. Each selected return receives the same displacement, independently of the size of its enclosed subset.

There is a larger version of the collective family: all six returns enclosing both other punctures can remain independently free. To see why, group the vacuum diagrams by these propagators. Each occurs at most once, and the coefficient of each is a lower-loop cut sum that vanishes on the deformed assignment; the part containing none vanishes as well. For a single-return displacement starting from Airy, the same cut isolates a sign-summed lower-loop cancellation. This explains why varying that return preserves the zero. Ordered external insertions and independent external-tree factors extend these identities to arbitrary multiplicity.

\paragraph{Additional finite families.}
For the low-point checks, choose $h_1=0$, $h_2=h$ and independent $A,B,h,h_3,U_i,V_i$. Begin with

\begin{equation}
 \begin{aligned}
 Y_{i,p}^+&=U_i+h_p,&Y_{i,p}^-&=V_i-h_p,\\
 Z_{pq}^{++}&=A+h_p+h_q,&Z_{pq}^{--}&=B-h_p-h_q,\\
 Z_{pq}^{+-}&=h_p-h_q,&Z_{pq}^{-+}&=h_q-h_p,\\
 T_{p,\{q\}}^+&=A+2h_p,&T_{p,\{q\}}^-&=B-2h_p.
 \end{aligned}
 \label{eq:three-low-internal}
\end{equation}

\begin{equation}
 X_{ij,S}=\begin{cases}
 U_i+V_j,&i=j\text{ or }1\in S,\\
 U_j+V_i,&i<j\text{ and }1\notin S.
 \end{cases}
 \label{eq:three-low-X}
\end{equation}

The connection formulas use $p<q$. The following completion also replaces $Y_{1,1}^+$ by an independent $w$:

\begin{equation}
 \begin{aligned}
 Z_{12}^{++}&=A+h+\epsilon,\\
 T_{1,S}^+&=A+\epsilon&&(2\in S),\\
 T_{2,S}^+&=A+2h+\epsilon&&(1\in S).
 \end{aligned}
 \label{eq:three-low-completion}
\end{equation}

The four other larger returns, $T_{1,\{2,3\}}^-$, $T_{2,\{1,3\}}^-$ and $T_{3,\{1,2\}}^\pm$, remain free.

At one point use two families. The first has no completion, $(U_1,V_1)=(U,V)$, and leaves all six larger returns and all three $X_{11,S}$ with $|S|=2$ independent; singleton $X$ values are $U+V$. The second uses the completion with $(U_1,V_1)=(0,-A)$, $w=A+h$ and $\epsilon=h$, so every $X=-A$. Include all puncture and sign images. Rational substitution proves both identities; together they reduce the 1008 coefficients to one.

At two points use the completion with $(U_1,V_1)=(0,-A)$, $(U_2,V_2)=(u,v)$ and $D_{1,2}=-u$. The parameters $w,\epsilon,D_{2,2}$ and the four larger returns are independent. Include both boundary placements and all puncture and sign images. With boundary zeros, this family fixes the two-point ansatz.

The following larger family is an alternative for $n\ge3$; it remains available in the package, while the default three-point check uses the six simpler families above:

\begin{equation}
 \begin{gathered}
 \begin{array}{c|cc}
 \text{mark}&U_i&V_i\\\hline
 R=1&0&-A\\
 I=2,\ldots,n-1&A+v_i&v_i\\
 J=n&A-h+v_n&v_n
 \end{array},\\[2pt]
 \epsilon=h,\quad w=A+h.
 \end{gathered}
 \label{eq:three-rij-profiles}
\end{equation}

Use Eqs.~\eqref{eq:three-low-internal}--\eqref{eq:three-low-completion}; every $v_i$, $D$ and the four larger returns are independent. Include cyclic placements and all puncture and sign images. The $R,I,J$ profiles coordinate the modified ray with the boundary curves. Substitution gives the three-point zero, including the sums obtained by deleting external trees. Splitting an $I$ mark repeats its profile and gives a divided difference in $v$, extending the identity to every $RI^{n-2}J$.

\subsection{Four loops}
\label{app:four}\label{sec:four-finite}\label{app:four-low-points}
We list three families at three points and the separate one- and two-point prescriptions. Their boundary insertions give the four-point starting zeros and the higher-point families. Let $P=\{1,2,3,4\}$, $h_1=0$, $h_2=h$, with $h_3,h_4$ independent.

\paragraph{Three-point families.}
Choose a distinguished puncture $r\in P$ and set

\begin{equation}
 \begin{aligned}
 Y_{i,p}^{+}&=U_i+h_p,&Y_{i,p}^{-}&=V_i-h_p,\\
 Z_{pq}^{++}&=A+h_p+h_q,&Z_{pq}^{--}&=B-h_p-h_q,\\
 Z_{pq}^{+-}&=h_p-h_q,&Z_{pq}^{-+}&=h_q-h_p,\\
 T_{p,\{q\}}^{+}&=A+2h_p,&T_{p,\{q\}}^{-}&=B-2h_p,
 \end{aligned}
 \label{eq:four-zero-family}
\end{equation}

\begin{equation}
 X_{ij,S}=\begin{cases}
 U_i+V_j,&i=j\text{ or }r\in S,\\
 U_j+V_i,&i<j\text{ and }r\notin S.
 \end{cases}
 \label{eq:four-zero-X}
\end{equation}

Connections use $p<q$ and boundary endpoints $i\le j$. All six $D$ and all 32 signed returns with $|S|\ge2$ are independent. Its relation to Airy is explicit under

\begin{equation}
 \begin{aligned}
 a_p&=A/2+h_p,& b_i&=(U_i+V_i)/2,\\
 \delta_i&=(U_i-V_i-A)/2,&\Delta&=A+B.
 \end{aligned}
 \label{eq:four-airy-parameters}
\end{equation}

Then $Y_{i,p}^{\pm}=b_i\pm(a_p+\delta_i)$, $Z_{pq}^{--}=-a_p-a_q+\Delta$ and $T_{p,\{q\}}^-=-2a_p+\Delta$. The other $Z$ channels and positive singleton returns keep their Airy values; the two $X$ values are $b_i+b_j\pm(\delta_i-\delta_j)$. Setting $\delta_i=\Delta=0$ and the larger returns to $2sa_p$ recovers Airy.

A second family deforms a pair $a<b$ and its associated returns:

\begin{equation}
 \begin{aligned}
 Z_{ab}^{++}&=A+h_a+h_b+\epsilon,\\
 T_{a,S}^{+}&=A+2h_a+\epsilon&& (b\in S),\\
 T_{b,S}^{+}&=A+2h_b+\epsilon&& (a\in S).
 \end{aligned}
 \label{eq:four-pair}
\end{equation}

The replacements include larger returns. A third family uses $r=1$ in the first assignment and frees one ray:

\begin{equation}
 \begin{aligned}
 (U_1,U_2,U_3)&=(u,A+v,A+t-h),\\
 (V_1,V_2,V_3)&=(-A,v,t),\qquad Y_{1,1}^{+}=w,
 \end{aligned}
 \label{eq:four-ray}
\end{equation}

\begin{equation}
 D_{1,3}=-(A+v),\qquad D_{1,2}=t-v.
 \label{eq:four-ray-caps}
\end{equation}

Keep $u$ in every other channel; all other $D$ and larger returns stay free. Take the first family with $r=1$, the pair $12$ deformation with $r=1$ or $3$, and the free-ray family. Include their puncture permutations and coherent sign exchanges.

\paragraph{One and two points.}
The following conditions supply the separate low-point checks. Use the internal assignment above and the pair deformation, with

\begin{equation}
 (A,h_2,h_3,h_4)=\lambda(7,2,5,11)
 \quad\text{or}\quad\lambda(11,3,7,17),
 \label{eq:four-low-ratios}
\end{equation}

where $\lambda\ne0$, while $B,\epsilon$ and the 26 unaffected larger signed returns are free. Include all puncture and sign images.

At one point deform pair $34$, take independent $U,V$, and use

\begin{equation}
 \begin{gathered}
 Y_{1,p}^+=U+h_p,\qquad Y_{1,p}^-=V-h_p,\\
 X_{11,S}=U+V\quad\text{if }|S\cap\{3,4\}|=1,
 \end{gathered}
 \label{eq:four-one-point-zero}
\end{equation}

The six other $X$ channels remain free. The constrained $X$ curves separate the deformed pair. Cutting these curves divides the diagram into lower-loop pieces; their cancellations permit the other six $X$ channels to remain free.

At two points deform pair $12$, set $C=\{3,4\}$, and use

\begin{equation}
 \begin{aligned}
 Y_{1,1}^+&=w,&Y_{1,p}^-&=-A-h_p,\\
 Y_{1,p}^+&=h_p+\tau\mathbf1_{p\in C}&&(p\ne1),\\
 Y_{2,p}^+&=A+v+h_p,&Y_{2,p}^-&=v-h_p.
 \end{aligned}
 \label{eq:four-two-point-rays}
\end{equation}

Here $\mathbf1$ is the membership indicator. Let $r(S)=1$ if $1\notin S$ and $S\cap C\ne\varnothing$, and zero otherwise. With $S^c=P\setminus S$, set

\begin{equation}
 \begin{aligned}
 X_{11,S}&=-A+\tau[r(S)+r(S^c)],\\
 X_{12,S}&=v+\tau r(S),\qquad X_{22,\{p\}}=A+2v,\\
 D_{1,2}&=-(A+v).
 \end{aligned}
 \label{eq:four-two-point-channels}
\end{equation}

The parameters $v,\tau,w,D_{2,2}$ and the ten nonsingleton $X_{22,S}$ are free. Include both boundary placements, complementing $S$ on $X_{12,S}$ when exchanging marks, and impose ordinary boundary zeros. Substitution and grouping by the free channels proves the zero. These one- and two-point families each leave one normalization in the local ansatz.

\paragraph{Starting checks and package coverage.}
The three-point four-loop families fix the full local ansatz there. At four points, the boundary equations leave the core attachment sums, whose triangular residues are the corresponding three-point core sums. Insert one leg into the three-point zeros using Appendix~\ref{app:boundary}. They fix these same weights and establish the four-point starting case. The subsequent induction is the one already described above.

The program-only WL package checks $n=1,2,3$ at two and three loops. It uses the two mountains throughout at two loops, the additional finite families at three-loop $n=1,2$, and precisely Eq.~\eqref{eq:three-six-family-list} at three-loop $n=3$. The four-loop calculations are separate. Every finite starting check leaves one overall normalization.

\end{document}